%% file: main.tex
\documentclass[%
twocolumn,
superscriptaddress,
longbibliography,
nofootinbib,
amsmath,amssymb,
aps,
notitlepage
]{revtex4-1}

\usepackage{graphicx}
\usepackage{dcolumn}
\usepackage{bm}
\usepackage{hyperref}
\usepackage{color}

\usepackage{amsmath}
\usepackage{acronym}
\usepackage{xspace}
\usepackage{subfigure}
\usepackage{longtable} 
\usepackage{multirow}
\usepackage{mathtools}
\usepackage{soul}

\usepackage{makecell}

\DeclarePairedDelimiter\abs{\lvert}{\rvert}

\newcommand{\orcidicon}[1]{\href{https://orcid.org/#1}{\includegraphics[height=\fontcharht\font`\B]{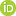}}}

\newcommand{\IUCAA}{Inter-University Centre for Astronomy and Astrophysics, Post Bag 4, Ganeshkhind, Pune 411 007, India}
\newcommand{\WSU}{Department of Physics \& Astronomy, Washington State University, 1245 Webster, Pullman, WA 99164-2814, USA}
\newcommand{\IPMU}{Kavli Institute for the Physics and Mathematics of the Universe (WPI), University of Tokyo, 5-1-5, Kashiwanoha, 2778583, Japan}
\newcommand{\UIB}{Departament de Física, Universitat de les Illes Balears, IAC3 - IEEC, Carretera de Valldemossa km 7.5, E-07122 Palma, Spain}
\newcommand{\AMU}{Aix-Marseille Université, Université de Toulon, CNRS, CPT, Marseille, France}

\newcommand{\kmsMpc}{\ensuremath{\mbox{km s}^{-1} \,\mbox{Mpc}^{-1}}\xspace}
\newcommand{\LCDM}{\ensuremath{\Lambda\mbox{CDM}}\xspace}

\newcommand{\bfr}{{\bf r}}
\newcommand{\bfx}{{\bf x}}
\newcommand{\xigwg}{\xi_{\rm gw, g}}
\newcommand{\xim}{\xi_{\rm m}}
\newcommand{\bgw}{b_{\rm gw}}
\newcommand{\bgal}{b_{\rm g}}
\newcommand{\ngal}{n_{\rm g}}

\newcommand{\nhalo}{n_{\rm h}}
\newcommand{\nhbar}{\bar{n}_{\rm h}}
\newcommand{\bhalo}{b_{\rm h}}

\begin{document}

\title{Bayesian framework to infer the Hubble constant from the cross-correlation of individual gravitational wave events with galaxies}

\author{Tathagata Ghosh~\orcidicon{0000-0001-9848-9905}}
\affiliation{\IUCAA}

\author{Surhud More~\orcidicon{0000-0002-2986-2371}}
\affiliation{\IUCAA} \affiliation{\IPMU}

\author{Sayantani Bera~\orcidicon{0000-0003-0907-6098}}
\affiliation{\AMU} \affiliation{\UIB} 

\author{Sukanta Bose~\orcidicon{0000-0002-4151-1347}}
\affiliation{\WSU} \affiliation {\IUCAA}

\begin{abstract}

Gravitational waves (GWs) from the inspiral of binary compact objects offer a one-step measurement of the luminosity distance to the event, which is essential for the measurement of the Hubble constant, $H_0$, which characterizes the expansion rate of the Universe. However, unlike binary neutron stars, the inspiral of binary black holes is not expected to be accompanied by electromagnetic radiation and a subsequent determination of its redshift. Consequently, independent redshift measurements of such GW events are necessary to measure $H_0$. In this study, we present a novel Bayesian approach to infer $H_0$ by measuring the overdensity of galaxies around individual binary black hole merger events in configuration space. We model the measured overdensity using the $3$D cross-correlation between galaxies and GW events, explicitly accounting for the GW event localization uncertainty. We demonstrate the efficacy of our method with $250$ simulated GW events distributed within $1$ Gpc in colored Gaussian noise of Advanced LIGO and Advanced Virgo detectors operating at O4 sensitivity. We show that such measurements can constrain the Hubble constant with a precision of $\lesssim 8 \%$ ($90\%$ highest density interval). We highlight the potential improvements that need to be accounted for in further studies before the method can be applied to real data.

\end{abstract}

\date{\today}

\maketitle

\section{Introduction}

The discovery of gravitational wave (GW) events opened an independent way to probe the expansion history of the Universe. The very first multimessenger approach to infer the Hubble constant was possible due to the detection of GW from the binary neutron star (BNS) merger, GW170817~\cite{LIGOScientific:2017vwq} along with the electromagnetic (EM) counterpart~\cite{LIGOScientific:2017ync}. 
The independent measurement of luminosity distance of the BNS merger from the GW data and its redshift from the EM data allows an inference of the Hubble constant $H_{0}=70^{+12.0}_{-8.0}$ \kmsMpc~\cite{LIGOScientific:2017adf}. 
However, most of the GW observations by the detectors in the LIGO-Virgo-KAGRA collaboration~\cite{LIGOScientific:2014pky, VIRGO:2014yos, KAGRA:2018plz} are primarily binary black hole (BBH) mergers~\cite{KAGRA:2021vkt}, which are not accompanied by EM counterparts. Such events are aptly called \textit{dark sirens}.
Though the unique identification of the host galaxy for such BBHs is not currently possible due to large sky localization errors, there are alternative approaches to infer the Hubble constant from such dark sirens.
The idea initially proposed by Schutz~\cite{1986Natur.323..310S} involves the use of galaxy catalogs for the identification of galaxies clustered within the sky localization of the GW events, which would provide noisy but useful information on the redshift of the events. His approach has been implemented as a statistical host identification method in a Bayesian framework (also sometimes termed the galaxy catalog method) by Refs.~\cite{DelPozzo:2011vcw, Chen:2017rfc, LIGOScientific:2018gmd, DES:2019ccw, Gray:2019ksv, LIGOScientific:2019zcs, LIGOScientific:2020zkf, Gray:2021sew, Finke:2021aom, Palmese:2021mjm, Gair:2022zsa, Alfradique:2023giv, DESI:2023fij}.
In another approach called the spectral siren method, the source redshift can be inferred by invoking the universality of the source frame mass distribution motivated from astrophysics~\cite{Chernoff:1993th, Taylor:2011fs, Farr:2019twy, You:2020wju, Mastrogiovanni:2021wsd, Mancarella:2021ecn, Ezquiaga:2022zkx}. Both of these methods, i.e., the galaxy catalog method and spectral siren method, have been applied to $47$ BBHs from GWTC-3~\cite{KAGRA:2021vkt} in addition to the bright siren GW170817 and its EM counterpart~\cite{LIGOScientific:2018hze}, to constrain the Hubble constant to $H_{0}=68^{+8}_{-6}$ \kmsMpc and $H_{0}=68^{+12}_{-8}$ \kmsMpc~\cite{LIGOScientific:2021aug}, respectively\footnote{The errors denote $68\%$ credible intervals in this case.}.
There are also ongoing efforts~\cite{Mastrogiovanni:2023emh, Gray:2023wgj, Stachurski:2023ntw, Borghi:2023opd} to unify the spectral and galaxy methods.
Furthermore, if at least one of the components of the merger is a neutron star, one can use the measurement of its tidal deformability and knowledge of the neutron star equation of state to infer the source redshift~\cite{Messenger:2011gi, Chatterjee:2021xrm, Jin:2022qnj, Ghosh:2022muc, Shiralilou:2022urk, Dhani:2022ulg, Ghosh:2024cwc}, and hence $H_{0}$ inference by combining the redshift information with the measured luminosity distance of the merger event.

The current state-of-the-art methods~\cite{Gray:2021sew, Mastrogiovanni:2021wsd} adopted to estimate $H_{0}$ in the recent 
LIGO-Virgo-KAGRA (LVK) cosmology analysis~\cite{LIGOScientific:2021aug} are dominated by the assumptions related to the nature of the GW source population~\cite{LIGOScientific:2021aug}. 
However, it is important to note that these two methods mentioned above do not utilize the information on the spatial clustering of galaxies explicitly. Both the GW source population and the galaxy population are tracers of the same underlying large-scale structure in the Universe. So, GW sources share the same large-scale structure as galaxies at their redshifts and, hence, are correlated to each other. Therefore, one can expect that the cross-correlation function between galaxies with known redshift and GW sources would be nonzero at the true redshift of the GW sources and, thus, in conjunction with the luminosity distances of the GW events, can be used to constrain the cosmological parameters such as the Hubble constant~\cite{MacLeod:2007jd, Oguri:2016dgk, Bera:2020jhx, Mukherjee:2019wcg, Mukherjee:2020hyn, Calore:2020bpd, Diaz:2021pem, Mukherjee:2022afz}.
In this work, we demonstrate a novel Bayesian framework to infer $H_{0}$ from \textit{individual} GW events by utilizing the cross-correlation between these GW events and the galaxy catalog. This approach models the galaxy overdensity for each GW event while also accounting for its volume uncertainty region when estimating the Hubble constant through this cross-correlation. This method, which can be termed as \textit{cross-correlation}, does not directly rely on assumptions related to the intrinsic population parameters that describe the GW sources.

The paper is structured as follows. In Sec.~\ref{sec:method}, we describe the methodology of using galaxy clustering information to estimate $H_{0}$ from individual GW events. After that, we elaborate on the simulations performed to demonstrate the methodology in Sec.~\ref{sec:simulation}. In Sec.~\ref{sec:results}, we show the efficacy of our method in recovering $H_{0}$ and examine how various parameters impact the precision of the resulting constraints. Finally, we conclude our paper and discuss future directions in Sec.~\ref{sec:summary}.

\section{Methodology} \label{sec:method}

\begin{figure*}
    \centering
    \includegraphics[scale=0.6]{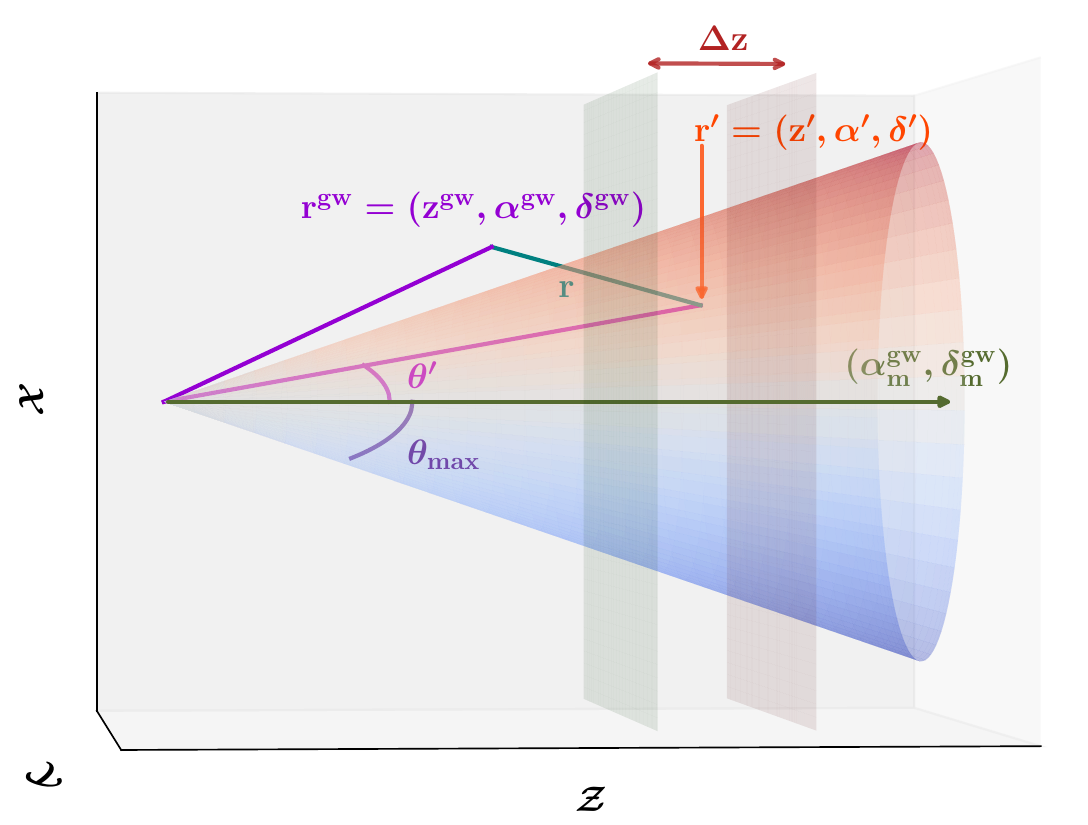}
    \caption{An illustration of the geometry in the three-dimensional Cartesian coordinate system $(\mathcal{XYZ})$ in order to compute the modeled galaxy overdensity $\delta_{g}^{\rm mod}$ following Eq.~\eqref{eq:del_mod}. 
    The horizontal arrow denotes the line of sight from the observer towards the maximum posterior probability position of the GW event ($\alpha^{\rm gw}_{\rm m}, \delta^{\rm gw}_{\rm m}$), the conical volume has an angular radius of $\theta_{\rm max}$ around this position, and the two vertical planes indicate the redshift range of width $\Delta z$, centered at $z$ within which we compute this modeled galaxy overdensity.}
    \label{fig:delmod_geometry}
\end{figure*}

In the standard model of cosmology (also known as the \LCDM model), the matter distribution in the Universe forms a cosmic web of large-scale structure with clustering properties determined by the value of cosmological parameters. Dark matter halos form at the peaks of the matter density distribution. The clustering properties of these halos on large scales reflect the clustering properties of the matter distribution, modified with a multiplicative factor $b_{h}^2$, where $b_{h}$ is the linear bias that corresponds to the ratio of the halo overdensity ($\delta_{\rm h}$) to the matter overdensity ($\delta_{\rm m}$),
\begin{align}
    \delta_{\rm h}(\bfx, M) \equiv \frac{\nhalo(\bfx, M)}{\nhbar}-1 &= \bhalo(M) \delta_{\rm m}(\bfx) \,.
\end{align}
Here, $\nhalo(\bfx, M)$ denotes the number of halos at a given position $\bfx$ with halo mass $M$ and $\nhbar$ is the average number of halos \cite{Cooray:2002dia}. The bias of the dark matter halos increases with increasing halo mass \citep{Kaiser:1984sw, Tinker_2010}. More luminous galaxies are expected to form within more massive dark matter halos \cite{Cacciato:2008hm, 2011ApJ...736...59Z, 2011MNRAS.410..210M, More:2011hz, 2019MNRAS.488.3143B}, and on large scales, they inherit the bias of the dark matter halos they inhabit, such that
\begin{align}
    \bgal = \int dM P_h(M) b_h(M)
\end{align}
where $P_h(M)$ specifies the distribution of halo masses to which the galaxies belong \cite{Sheth:1999mn, Tinker_2010, Lazeyras:2016xfh} (for a detailed review on the halo and galaxy bias, one may refer to \cite{DESJACQUES20181}).

In this section, we describe the methodology to utilize the cross-correlation between individual GW events and galaxy catalogs to constrain the Hubble constant. 
The spatial clustering between galaxies and GW events, separated by the comoving distance $\bfr$, can be characterized by the $3$D cross-correlation function, $\xi_{\rm gw, g}(\bfr)$, given by
\begin{align} \label{eq:xi_def}
    \xi_{\rm gw, g} (\bfr) \equiv \langle\delta_{\rm gw}(\bfx) \delta_{\rm g}(\bfx + \bfr)\rangle \,,
\end{align}
where $\bfx$ represents the location of GW sources; $\delta_{\rm gw}$ and $\delta_{\rm g}$ denote the overdensity of the GW source distribution and the galaxy distribution, respectively. As the Universe is assumed to be isotropic, the cross-correlation function depends on the scalar distance $r=\abs{\bfr}$.
Since GW events and galaxies are tracers of the same underlying matter distribution, both GW overdensity and galaxy overdensity are related to the matter overdensity $\delta_{m}$ on large scales as
\begin{subequations} \label{eq:bias}
\begin{gather}
    \delta_{\rm g} = b_{\rm g}\delta_{m} \\
    \delta_{\rm gw} = b_{\rm gw}\delta_{m}
\end{gather}
\end{subequations}
Here, the proportionality constants $b_{\rm g}$ and $b_{\rm gw}$ are the linear biases of the population of galaxies and GW sources with respect to the matter distribution, respectively. A bias of unity would imply that the observed population follows the underlying matter distribution. In general, the bias parameters depend on different properties of the sources. The dependence of galaxy bias on redshift can be estimated from the measurement of autocorrelation functions of galaxies in a given redshift slice with themselves. The redshift-dependent GW bias needs to be assumed a parametric form and be marginalized over those parameters (see Ref.~\cite{Mukherjee:2020hyn}).

The cross-correlation function $\xigwg$, defined in Eq.~\eqref{eq:xi_def} can be written as,
\begin{align}
    \xigwg = \bgw \bgal \xim\,.
\end{align}
Here, $\xim$ is the nonlinear matter power spectrum~\cite{Takahashi:2012em}, which we compute under the assumption of the standard \LCDM model of cosmology. 
In the flat \LCDM Universe, the luminosity distance $d_{L}$ is related to the redshift $z$ as follows:
\begin{align} \label{eq:z_dl_relation}
    d_{L} &= \frac{c(1+z)}{H_{0}} \int_{0}^{z} \frac{dz^{\prime}}{H(z^{\prime})}\nonumber\\
          &= d_{c}(1+z)
\end{align}
Here, $d_{c}$ denotes the comoving distance, and $H(z)$ is the Hubble parameter, expressed in terms of the Hubble constant $H_{0}$ and matter density parameter $\Omega_{m}$ as:
\begin{align} \label{eq:hubble_parameter}
    H(z)= H_{0} E(z) = H_{0} \sqrt{\Omega_{m}(1+z)^{3}+(1-\Omega_{m})}\ .
\end{align}

Since the correlation function $\xigwg(r)$ quantifies the excess
probability over a random Poisson distribution of finding a galaxy in a volume element $dV$, which is separated by a comoving distance $\bfr$ from a GW event located at position $\bfx$, we have
\begin{align} \label{eq:2point_correlation}
    \ngal(\bfx+\bfr) dV =\bar{n}_{V}\left[1+\xigwg(r)\right]dV\,,
\end{align}
where $\bar{n}_{V}$ is the average number density of galaxies in comoving volume.
So, the galaxy overdensity can be calculated by integrating Eq.~\eqref{eq:2point_correlation} over some finite volume if we have the theoretical prediction of the nonlinear matter power spectrum $\xim$.

Consider a GW event located at its true position ${\bf r}^{\rm gw}$ denoted by redshift $z^{\rm gw}=z^{\rm gw}(d_{L}^{\rm gw}, H_{0})$, right ascension $\alpha^{\rm gw}$, and declination $\delta^{\rm gw}$. The modeled galaxy overdensity, $\delta_{g}^{\rm mod}$, in an angle $\theta_{\rm max}$ around the maximum \textit{a posteriori} probability (MAP) location ($\alpha^{\rm gw}_{\rm m}, \delta^{\rm gw}_{\rm m}$) and within a redshift bin centered around $z^{\prime}$, can be computed by carrying out the following volume integral (see Fig.~\ref{fig:delmod_geometry} for the geometry) assuming the MAP is along the line of sight:
%
%
\begin{widetext}
\begin{align} \label{eq:del_mod}
\delta_{g}^{\rm mod}\left(z \mid {\bf r}^{\rm gw} \right) = \frac{\int_{z-\Delta z/2}^{z+\Delta z/2} \int_{0}^{\theta_{\rm{max}}} \int_{0}^{2\pi} \bar n_{V}(z^{\prime}) \xi_{\rm gw, g}(|{\bf r} ({\bf r}^{\rm gw}, {\bf r}^{\prime})|) \sin\theta^{\prime} d_{c}^{\prime~2} J\left(\frac{d_{c}^{\prime}}{z^{\prime}}\right) d\alpha^{\prime} d\theta^{\prime} dz^{\prime}}{\int_{z-\Delta z/2}^{z+\Delta z/2} \int_{0}^{\theta_{\rm{max}}} \int_{0}^{2\pi} \bar n_{V}(z^{\prime}) \sin\theta^{\prime} d_{c}^{\prime~2} J\left(\frac{d_{c}^{\prime}}{z^{\prime}}\right) d\alpha^{\prime} d\theta^{\prime} dz^{\prime}}\ .
\end{align}
\end{widetext}
Here, $\bf{r}$ denotes the comoving distance (in $h^{-1}$ unit) from ${\bf r}^{\rm gw}$ to a position at ${\bf r^{\prime}}=(z^{\prime}, \alpha^{\prime}, \delta^{\prime})$. Without loss of generality, we define our coordinate system such that $\delta_{\rm m}^{\rm gw}=0$. For any other orientation, we can apply an appropriate rotation of the coordinate axis. Based on this geometry, in Eq.~\eqref{eq:del_mod}, we use $\theta^{\prime}=\pi/2-\delta^{\prime}$.

The true GW event redshift can be inferred from the true luminosity distance by inverting Eq.~\eqref{eq:z_dl_relation}. The quantity $d_{c}^{\prime}$ in the above equation denotes the comoving distance corresponding to a redshift $z^{\prime}$ within the redshift bin enclosed by the vertical planes and $J\left(\frac{d_{c}^{\prime}}{z^{\prime}}\right)$ corresponds to the Jacobian of the coordinate transformation from comoving distance to redshift, which can be written as [see Eq.~\eqref{eq:z_dl_relation}]
\begin{align} \label{eq:jacobian}
    J\left(\frac{d_{c}}{z}\right)=\frac{\partial d_{c}}{\partial z}=\frac{c}{H(z)}\,.
\end{align}
The modeled galaxy overdensity thus depends on the choice of $H_{0}$.

The \textit{observed} galaxy overdensity within the redshift bin, centered around $z$, can be measured within the angular distance $\theta_{\rm max}$ around the line of sight towards the MAP position ($\alpha^{\rm gw}_{\rm m}, \delta^{\rm gw}_{\rm m}$), as
\begin{align} \label{eq:del_obs}
    \delta_{g}^{\rm obs}(z) = \frac{n(z \mid \delta^{\rm gw}_{\rm m},\alpha^{\rm gw}_{\rm m})}{\bar{n}(z)} - 1\,,
\end{align}
where $n(z \mid \delta^{\rm gw}_{\rm m},\alpha^{\rm gw}_{\rm m})$ denotes the number of galaxies within the redshift bin. For a galaxy survey with uniform depth, the average number of galaxies $\bar{n}(z)$ expected within the redshift bin can be calculated by averaging the observed number of galaxies within the same angular radius $\theta_{\rm max}$ around random lines of sight. 

The observed overdensity $\delta_{g}^{\rm obs}$ can be calculated in different redshift bins of galaxies. We call this the galaxy data vector $\bm{d}_{g}^{\rm obs}$. This galaxy data vector can be compared with the model vector of $\delta_{g}^{\rm mod}$ for the same redshift bins of galaxies, denoted by $\bm{d}_g^{\rm mod}$ to construct a galaxy overdensity likelihood. Note that the model vector is conditioned on the true location (both on the sky and in redshift) of a particular GW event. The comparison of the two should provide a constraint on the Hubble constant, $H_0$.

The true position on which the modeled galaxy overdensity is conditioned has its own uncertainty as the sky localization of the GW event is not precisely known but known as a posterior distribution over $\bm{d}_{\rm gw}=\{z^{\rm gw}(d_{L}^{\rm gw}, H_{0}), \alpha^{\rm gw}, \delta^{\rm gw}\}$ inferred from the gravitational wave strain data $\bm{d}_{\rm strain}$. The posterior for the Hubble constant, in that case, is given by,
\begin{widetext}
\begin{align} \label{eq:posterior_H0}
    p(H_{0} \mid \bm{d}_{\rm strain}, \bm{d}_{g}^{\rm obs}) &= \int p(H_0, \bm{d}_{\rm gw} \mid \bm{d}_{\rm strain}, \bm{d}_{g}^{\rm obs}) d \bm{d}_{\rm gw} \nonumber \\ 
    &\propto \int {\cal L}(\bm{d}_{\rm strain}, \bm{d}_{\rm g}^{\rm obs}|H_{0}, \bm{d}_{\rm gw}) P(H_0, \bm{d}_{\rm gw}) d\bm{d}_{\rm gw} \nonumber \\
    &\propto \int {\cal L}(\bm{d}_{\rm g}^{\rm obs}|H_{0}, \bm{d}_{\rm gw}) {\cal L}(\bm{d}_{\rm strain}|H_{0}, \bm{d}_{\rm gw}) P(H_0, \bm{d}_{\rm gw}) d\bm{d}_{\rm gw}\,,
\end{align}
\end{widetext}
where we have applied Bayes' theorem and assumed independence of the galaxy overdensity and the strain data from gravitational wave events. The integration in the last equation is performed over the localization uncertainty of the GW event. The first likelihood in Eq.~\eqref{eq:posterior_H0} corresponds to the galaxy overdensity likelihood, and we assume it to be a Gaussian such that
\begin{widetext}
\begin{align} \label{eq:likelihood_g}
    \mathcal{L} (\bm{d}_{g}^{\rm obs} \mid H_0, \bm{d}_{\rm gw}) \propto \exp\left[-\frac{1}{2} \left(\bm{d}_{g}^{\rm obs}-\bm{d}_{g}^{\rm mod}\right)^{\rm T} \bm{C}^{-1} \left(\bm{d}_{g}^{\rm obs}-\bm{d}_{g}^{\rm mod}\right) \right]
\end{align}
\end{widetext}
Here, $\bm{C}$ denotes the covariance of the overdensity field, and we assume it to be diagonal with individual elements $\sigma^2_{\delta}$ corresponding to the variance in the observed galaxy overdensity in a given redshift bin within the angle $\theta_{\rm max}$. The procedure for calculating $\sigma_{\delta}$ using random lines of sight is described in Sec.~\ref{sec:simulation}.

The second likelihood term in the last integral of Eq.~\eqref{eq:posterior_H0} corresponds to the strain data. Often, the posterior distributions of $P(\bm{d}_{\rm gw})$ are computed assuming certain priors. Therefore, the likelihood can be obtained by taking the ratio of the posteriors for source parameters in $\bm{d}_{\rm gw}$ and the priors that are used to infer these parameters. This allows us to separately impose our own priors on the quantities $P(H_{0}, \bm{d}_{\rm gw})$ in  Eq.~\eqref{eq:posterior_H0}. 

Since GW events are independent, the final joint posterior of the Hubble constant can be obtained by combining the independent likelihoods of all the detected sources,
\begin{widetext}
\begin{align} \label{eq:pH0_combine}
    P(H_{0} \mid \{\bm{d}_{\rm strain}\}, \{\bm{d}_{g}^{\rm obs}\}) &\propto P(H_{0}) \prod_{i} {\cal L}(\bm{d}_{{\rm strain}_{i}}, \bm{d}_{g_{i}}^{\rm obs}| H_0) \nonumber  \\
    &\propto P(H_{0}) \prod_{i} \int {\cal L}(\bm{d}_{{\rm strain}_{i}}, \bm{d}_{g_{i}}^{\rm obs}| H_0, \bm{d}_{\rm gw}) P(\bm{d}_{\rm gw}) d\bm{d}_{\rm gw}
\end{align}
\end{widetext}
where $P(H_{0})$ represents the prior over $H_{0}$,
the index $i$ denotes individual GW events, and $\bm{d}_{g_{i}}^{\rm obs}$ is the galaxy overdensity data vector corresponding to the $i{\rm th}$ GW event.

It is important to highlight that although the Bayesian formalism proposed in this work may look structurally similar to population inference techniques using GWs ~\cite{LIGOScientific:2020kqk, KAGRA:2021duu} and other cosmological inference methods such as spectral siren method and galaxy catalog method, it does not share the same selection biases as mentioned in Ref.~\cite{Mandel:2018mve}. Unlike those methods, the present work does not explicitly rely on mass-model assumptions. Rather, it measures and models observed galaxy overdensity associated with the detected GW events, determined by an SNR threshold. The observable here is the galaxy overdensity around the location of GW events, which we have assumed to be independent of the population properties of the GW events. Eq. \eqref{eq:pH0_combine} signifies that we are only taking into account the detected events and not using the events that are below our SNR threshold. While this introduces a bias for population inference, our method is expected to be free from such biases, since the overdensity maps (or the 3-D power spectrum) of GWs, for both detected or undetected events, are the same by our construction of the mock events. Hence we are not introducing any bias by leaving out subthreshold events. Even in the case that there are correlations between GW event properties and their host galaxy properties, those will be absorbed by the GW overdensity bias, $b_{GW}$, and its redshift evolution. For our simulated catalog, we have currently set it to unity. Similarly, the effect of flux limited galaxy surveys will have to be accounted for through the galaxy bias term $b_g$. We plan to examine such effects and their impact on cosmological inference in future work.

When constructing the posterior in Eq.~\eqref{eq:posterior_H0}, we take the flat \LCDM cosmology to be the true cosmological model, with $H_{0}$ 
as a free parameter and the matter density $\Omega_{m}$ kept fixed at $0.307$. In the flat \LCDM Universe, the relation between luminosity distance $d_{L}$ and the redshift $z$ in Eq.~\eqref{eq:z_dl_relation} is used to convert the measured luminosity distance of each GW event to its redshift for a given $H_0$ whenever required while performing the integral over $\bm{d}_{\rm gw}$ in Eq.~\eqref{eq:posterior_H0}.

We would like to emphasize here that, although this work stems from the same core idea (i.e., leveraging the clustering of GW events with other large-scale structure tracers) as outlined in Refs.~\cite{Bera:2020jhx, 1986Natur.323..310S}, the methodology presented here is not only significantly different but also incorporates several improvements over the previous work. As an example, consider Ref.~\cite{Bera:2020jhx}, which uses the angular cross-correlation between GW events and galaxies in different redshift bins to infer the Hubble constant in two steps.
In the first step, the GW events are binned in multiple luminosity distance bins based on the median luminosity distance estimate from the posterior distribution of parameters given the GW data. The GW events in each bin are cross-correlated with galaxies in narrow redshift bins in order to estimate the mean redshift of the GW events in that bin. The peak of the angular cross-correlation technique was used to infer the luminosity distance-redshift diagram. In the second step, the luminosity distance-redshift relation thus obtained was fit in the context of an assumed model ($\Lambda$CDM in this case) to constrain the Hubble constant using the standard Bayesian inference technique. In contrast, the method we present in this work involves modeling the galaxy overdensity for each GW event while accounting for its volume uncertainty region and estimating the Hubble constant by modeling the $3$D cross-correlation. We provide a detailed comparison of these two methodologies in the Appendix.

\section{Simulation} \label{sec:simulation}

\begin{figure*}
    \centering
    \includegraphics[scale=0.44]{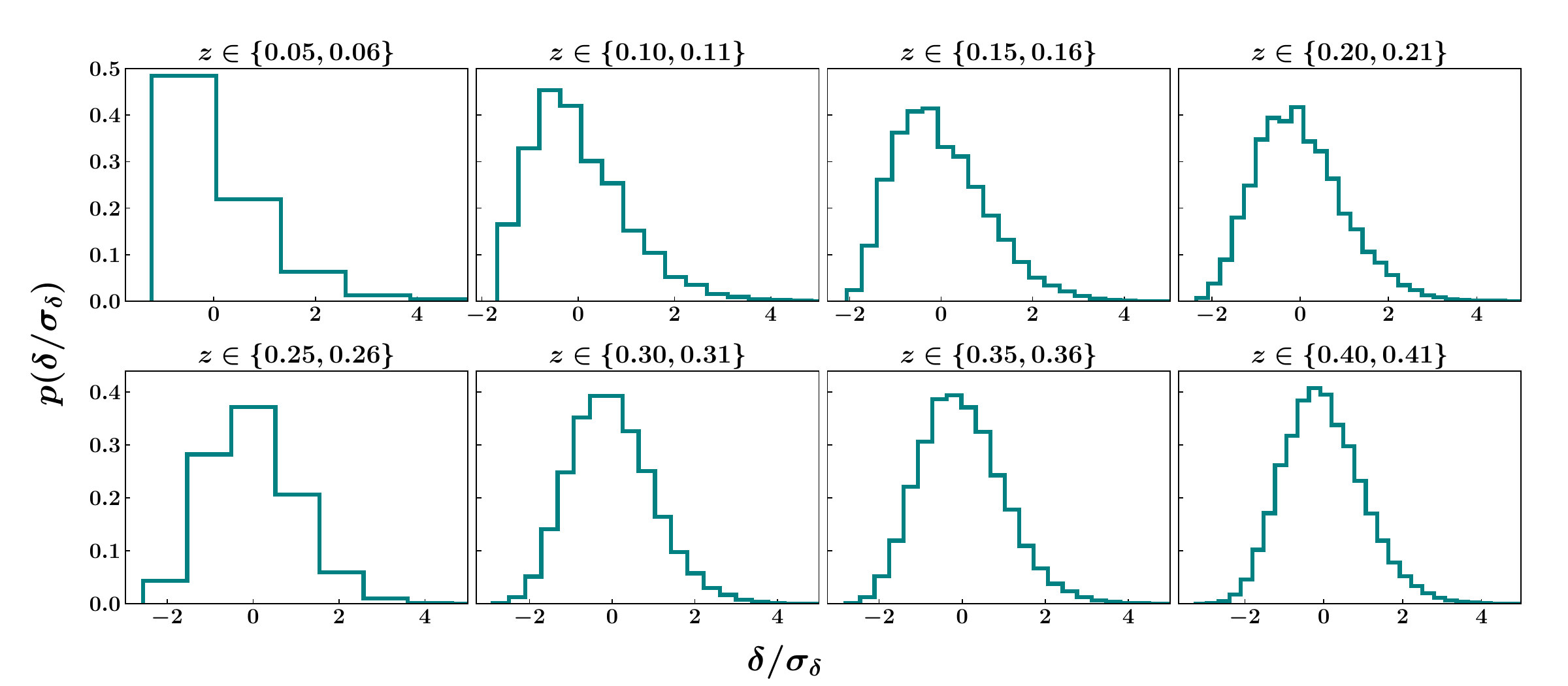}
    \caption{Probability distribution of observed overdensity divided by the corresponding fluctuation in different redshift bins, mentioned at the top of each panel, considering $\theta_{\rm max}=0.03$ radian.}
    \label{fig:delobs_hist}
\end{figure*}

We test the framework discussed in Sec.~\ref{sec:method} with a mock galaxy and GW catalog in order to assess how well the expansion rate can be recovered given realistic localization errors in the positions of GW events.
We consider dark matter halos from the Big MultiDark Planck (BigMDPL) cosmological N-body simulation~\cite{Klypin:2014kpa} from the CosmoSim database~\footnote{This is publicly available at \url{https://www.cosmosim.org/}.} to construct the galaxy catalog.
The BigMDPL simulation follows the evolution of the matter distribution in a flat \LCDM cosmology with the following parameters: Hubble parameter $h=H_{0}/(100\ \kmsMpc )=0.6777$, the matter density parameter $\Omega_{m}=0.307$, the amplitude of density fluctuations characterized by $\sigma_{8}=0.823$, and the power spectrum slope of initial density fluctuations $n_{s}=0.96$. The simulation is set up as $3840^{3}$ collisionless particles in a cubic box with comoving side length $2.5 h^{-1}{\rm Gpc}$ with periodic boundary conditions and corresponds to a mass resolution of $2.359\times 10^{10} h^{-1} M_{\odot}$.

We only consider well-resolved dark matter halos at the $z=0$ snapshot of BigMDPL,
with masses $\geq M_{\rm th}=10^{12} h^{-1}M_{\odot}$. For simplicity, we assume that all galaxies are central galaxies and thus place each galaxy at the center of these dark matter halos. We place an observer at the center of the box and compute the sky positions of each of these mock galaxies with respect to the center, which is within the sphere of radius $1.25 h^{-1}{\rm Gpc}$. We utilize the cosmological parameters of the simulation to determine the redshift of each galaxy from its comoving distance from the central observer. We use this mock galaxy catalog for further analysis. For simplicity, we have ignored any redshift space distortion effects.
In practice, galaxy catalogs obtained from real galaxy surveys suffer from incompleteness issues due to the sky-position dependent flux limits. Flux-limited galaxy catalogs must be considered to understand the effect of the limited number of observed fainter galaxies at higher redshift. In such cases, one has to account for how the flux limit introduces an evolution in the galaxy bias factor. This is expected to broaden the $H_{0}$ posterior compared to that obtained using a complete galaxy catalog. Similar effects are also expected for galaxy catalogs that have limited sky coverage. For such cases, the effect of appropriate sky masks needs to be considered while modeling the measurements. For simplicity, we consider a complete galaxy catalog with full sky coverage in this study.

Once we have the mock galaxy catalog, we randomly subselect galaxies to be the hosts of GW events without any regard to the properties of these halos or, equivalently, the properties of the galaxies in the catalog. The true redshifts and sky positions of the GW events are determined from their respective host galaxies in the cosmological box. We use the same cosmological parameters of the simulation box in order to compute the luminosity distance of GW events.
The galaxy catalog we use extends up to a comoving distance of $1.25 h^{-1}{\rm Gpc}$ given the size of the box before we run into incompleteness issues. This comoving distance corresponds to redshift $z\sim 0.46$, assuming the true cosmology. In this work, however, we consider uniform prior over $H_{0}$ between $20$ and $120~\kmsMpc$. Consequently, a higher value of $H_{0}$ can lead to an increase in the redshift of a GW event at a given luminosity distance.
Thus, the GW event can have redshift up to $z \sim 0.36$ when $H_{0}=120~\kmsMpc$ and yet be detectable in the box.
Therefore, we restrict the injection of GW events within a luminosity distance of $1$ Gpc to avoid any biases creeping in due to the above selection effects.

The distribution of the source-frame masses of the BBHs follows the Power Law+Peak model~\cite{Talbot:2018cva} between minimum mass $m_{\rm min} = 5 M_{\odot}$ and maximum mass $m_{\rm max} = 60 M_{\odot}$. The injected values of other mass hyperparameters (see Appendix B of Ref.~\cite{LIGOScientific:2020kqk} for the descriptions of the hyperparameters) are $\lambda_{\rm peak}=0.06, \alpha=3.5, \delta_{m}=4, \mu_{m}=34, \sigma_{m}=5$, and $\beta=1.1$, which are consistent with the previous observations~\cite{LIGOScientific:2020kqk, KAGRA:2021duu}.
However, note that the detected masses are redshifted and are expected to differ from the source-frame mass, $m$, as $m_{z}=m(1+z)$, where  $z$ is the redshift of the event. The other BBH parameters (see Table 1 of Ref.~\cite{Ashton_2019} for the list of parameters) are randomly chosen from the default priors of ~\verb+bilby+ (see Table 1 of Ref.~\cite{Ashton_2019} for the default priors).  
The simulated BBH signal is added to stationary Gaussian noise, colored with the O4 design sensitivity~\cite{KAGRA:2013rdx} of LIGO and Virgo detectors. We analyze the strain signal between $20$ and $2048$ Hz to infer the source parameters.
We consider the same nonprecessing waveform model~\verb+IMRPhenomD+~\cite{Khan:2015jqa} for both the signal injection and recovery~\footnote{Note that we consider these simplified waveform model, as we would like to demonstrate a proof-of-concept for the cross-correlation based framework that we present in this paper. Analysis of real events will involve upgraded waveform models, potentially including higher-order modes~\cite{London:2017bcn} and precession~\cite{Vitale:2018wlg}, which can better constrain the luminosity distance by breaking the degeneracy between luminosity distance and inclination angle. In this context, the $H_{0}$ measurement can be regarded as a conservative estimate for the simulated events considered in this work.}. We have used the nested sampler~\verb+dynesty+~\cite{2020MNRAS.493.3132S} implemented in~\verb+bilby+~\cite{Ashton_2019} to infer the posterior distribution of all the parameters given the waveform.
We consider uniform prior over observed chirp mass within the range $\mathcal{M}_{c, \text{inj}}^{z} \pm 5 \ M_{\odot}$, with the constraint that the component mass remains above $2.5~M_{\odot}$ and over mass-ratio $q=m_{2}/m_{1} \in [0.001, 1]$.
These choices of mass parameters and their specified prior ranges optimize sampling efficiency while lowering computational costs. 
We employ cosmology independent $d_{L}^{2}$ prior to the luminosity distance of the source between $10$ Mpc and $5$ Gpc.
The priors over the rest of the parameters (except for the mass parameters and luminosity distance) are the default priors implemented in~\verb+bilby+ (refer to Table $1$ in Ref.~\cite{Ashton_2019} for the default priors). 

\begin{figure}
    \centering
    \includegraphics[scale=0.42]{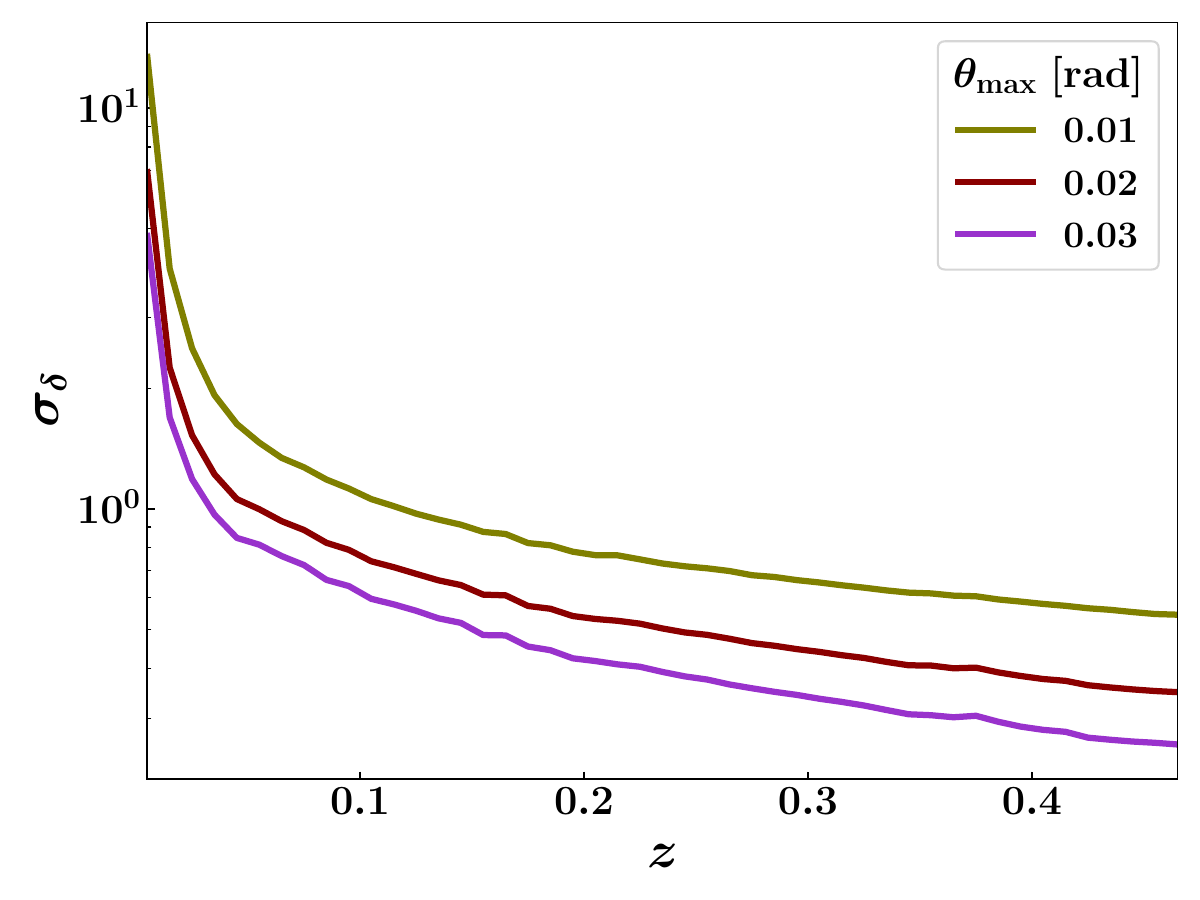}
    \caption{Observed overdensity fluctuation as a function of redshift over $10^{6}$ random line of sights.}
    \label{fig:sigma}
\end{figure}

The parameters thus inferred are utilized to constrain $H_{0}$ by cross-correlating the localization of these individual events with galaxies in the mock catalog. We are primarily interested in the extrinsic source parameters, such as $d_{L}^{\rm gw}$, $\alpha^{\rm gw}$, and $\delta^{\rm gw}$ in this work. So, we compute the GW likelihood $\mathcal{L}_{\rm gw}$ by marginalizing over the rest of the intrinsic and extrinsic parameters. The galaxy overdensity likelihood $\mathcal{L}_{g}$ is also calculated for different sky coordinates within the localization of the GW events as a function of redshift in order to perform the integration of Eq.~\eqref{eq:posterior_H0}. 
The galaxy overdensity likelihood includes the computation of the observed and modeled galaxy overdensity fields as a function of redshift. 
The modeled galaxy overdensity is calculated from the cross-correlation function $\xi$ (see, Eq.~\eqref{eq:del_mod}), which depends on the separation (comoving distance in $h^{-1}$ scale) between the GW event and the redshift where the observed galaxy overdensity is also measured. 
We have used the publicly available code \verb+AUM+~\footnote{\url{http://surhudm.github.io/aum/index.html}} to compute $\xi$~\cite{2013MNRAS.430..725V} from the nonlinear matter power spectrum~\cite{Takahashi:2012em} assuming the flat \LCDM cosmology.
To study the effect of the cross-correlation length scale on the $H_{0}$ estimation, we assume three different values of the angular radius $\theta_{\rm max}$ (elaborated in Sec.~\ref{sec:results}) along the line of sight to calculate the observed and modeled galaxy overdensity fields.

In order to compute the likelihood of observing the galaxy overdensity given the modeled overdensity, we quantify the variance of the galaxy overdensity field [denoted as $\sigma^2_{\delta}$ in Eq.~\eqref{eq:posterior_H0}] around random lines of sight at each redshift bin. The variance of the overdensity depends upon the specific value of $\theta_{\rm max}$ over which we average. We use $10^{6}$ random lines of sight within the mock galaxy catalog used in this work. The probability distribution of $\delta/\sigma$\footnote{We are omitting superscripts and subscripts.} for eight representative redshift bins of width $\Delta z=0.01$ up to $z=0.41$ are shown in Fig.~\ref{fig:delobs_hist}, for $\theta_{\rm max}=0.03$ radian.
The figure clearly shows that $\delta/\sigma$ becomes non-Gaussian at lower redshifts. Despite this, we assume a Gaussian likelihood for galaxy overdensity for simplicity and find that this assumption does not affect $H_{0}$ estimation in this study, given the sensitivity of the GW detectors considered. Developing a more rigorous framework for modeling the galaxy overdensity likelihood can be explored in future work.
The corresponding standard deviation of the overdensity field is shown as a function of redshift as solid lines in Fig.~\ref{fig:sigma}, where the colors signify different values of $\theta_{\rm max}$. As expected, $\sigma$ decreases with redshift for any $\theta_{\rm max}$ due to the increase in cosmological volume probed within the angular region per redshift interval. 
Similarly, a higher value of $\theta_{\rm max}$ also results in a smaller standard deviation for the overdensity at any given redshift. 
Notably, the redshift width of $\Delta z=0.01$ for calculating galaxy overdensity has been carefully selected to ensure a sufficient number of galaxies in each redshift bin.
We empirically choose the optimal value of $\Delta z=0.01$ by experimenting with various choices, optimizing the impact of Poisson noise on the overdensity measurement for smaller $\Delta z$ and the diminished cross-correlation signal for larger $\Delta z$. In a realistic scenario, this choice needs to be adjusted depending on the galaxy number densities used for the cross-correlation.
We use this variance in order to evaluate the likelihood of the galaxy overdensity given the modeled overdensity and a true position of the GW event, which, among other parameters, depends upon the Hubble constant. In order to infer the posterior distribution of the Hubble constant given the data, we assume a uniform prior in $H_{0} \in U [20, 120]\ \kmsMpc$. We finally infer the joint posterior for $H_{0}$ for multiple events as given by Eq.~\eqref{eq:pH0_combine}.

\section{Results} \label{sec:results}

\begin{figure*}
    \centering
    \includegraphics[scale=0.39]{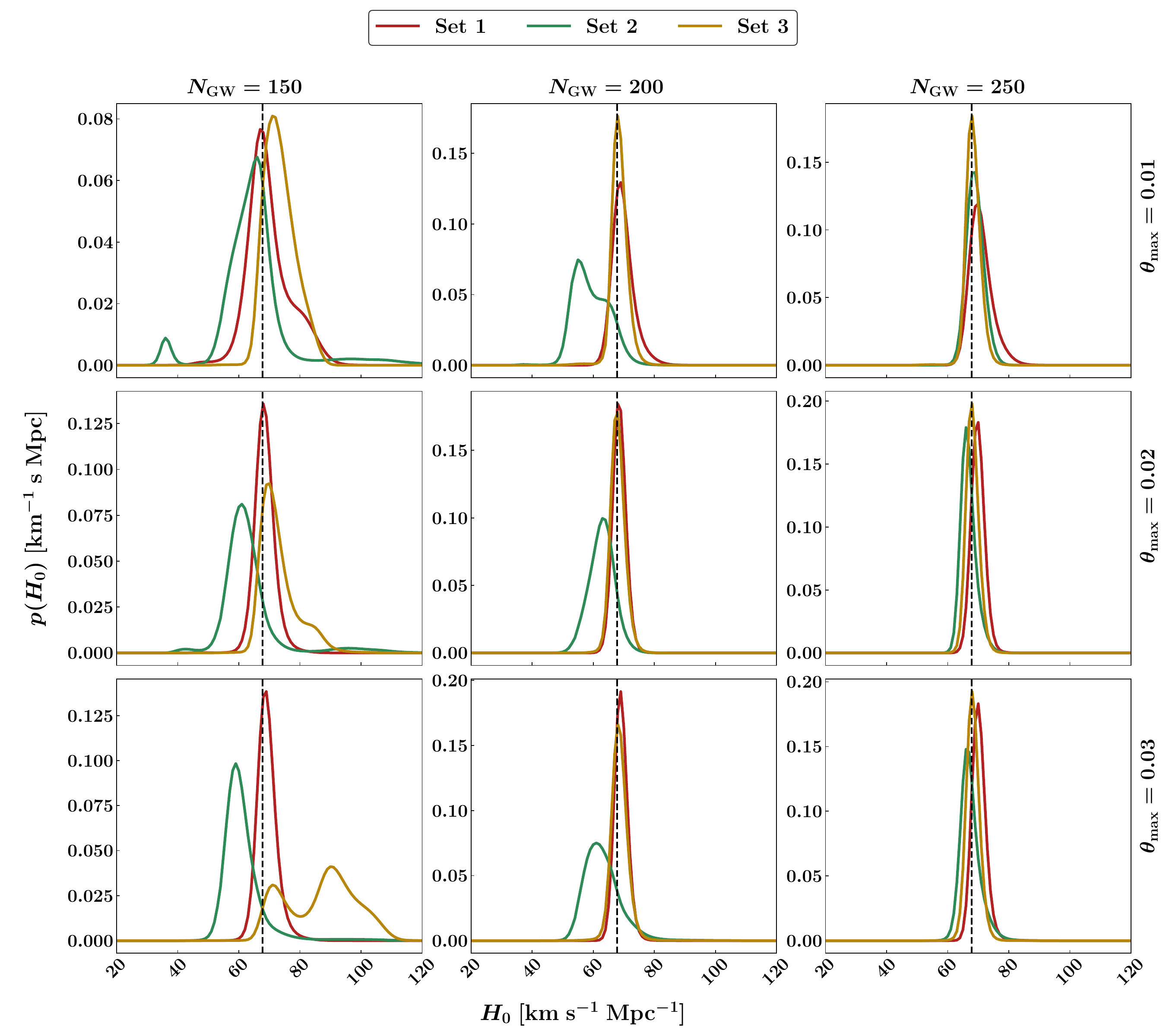}
    \caption{Estimation of Hubble constant from BBH events whose network SNR $\rho_{\rm net} \geq 12$, as detected by LIGO-Virgo detector operating at O4 sensitivity, for different choices of $\theta_{\rm max}$ (in radian). The black vertical dashed line in each plot corresponds to the injected value of $H_{0}= 67.77~\kmsMpc$.}
    \label{fig:H0_snr12}
\end{figure*}

\begin{figure}
    \centering
    \includegraphics[scale=0.4]{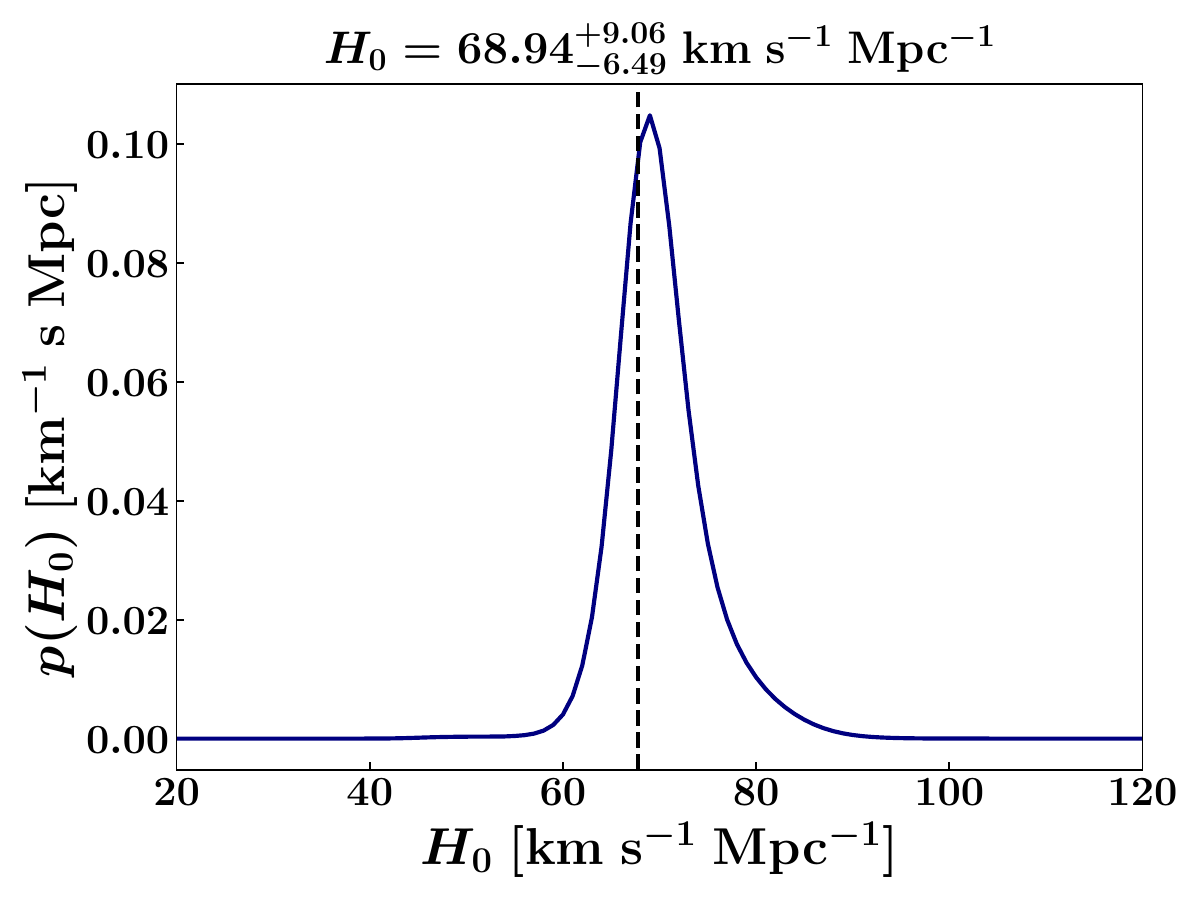}
    \caption{Posterior distribution of the Hubble constant inferred using $\theta_{\rm max}=0.02$ radian from one realization of $50$ randomly chosen BBH events, each with a network SNR $\rho_{\rm net} \geq 12$, as detected by the LIGO-Virgo detectors operating at O4 sensitivity. The accuracy with which $H_0$ is inferred in this realization is representative amongst the numerous random realizations we constructed. The median and $90\%$ highest density interval are indicated at the top of the plot. The black vertical dashed line in the plot represents the injected value of $H_{0}= 67.77~\kmsMpc$.}
    \label{fig:H0_rand50}
\end{figure}

\begin{figure*}
    \centering
    \includegraphics[scale=0.44]{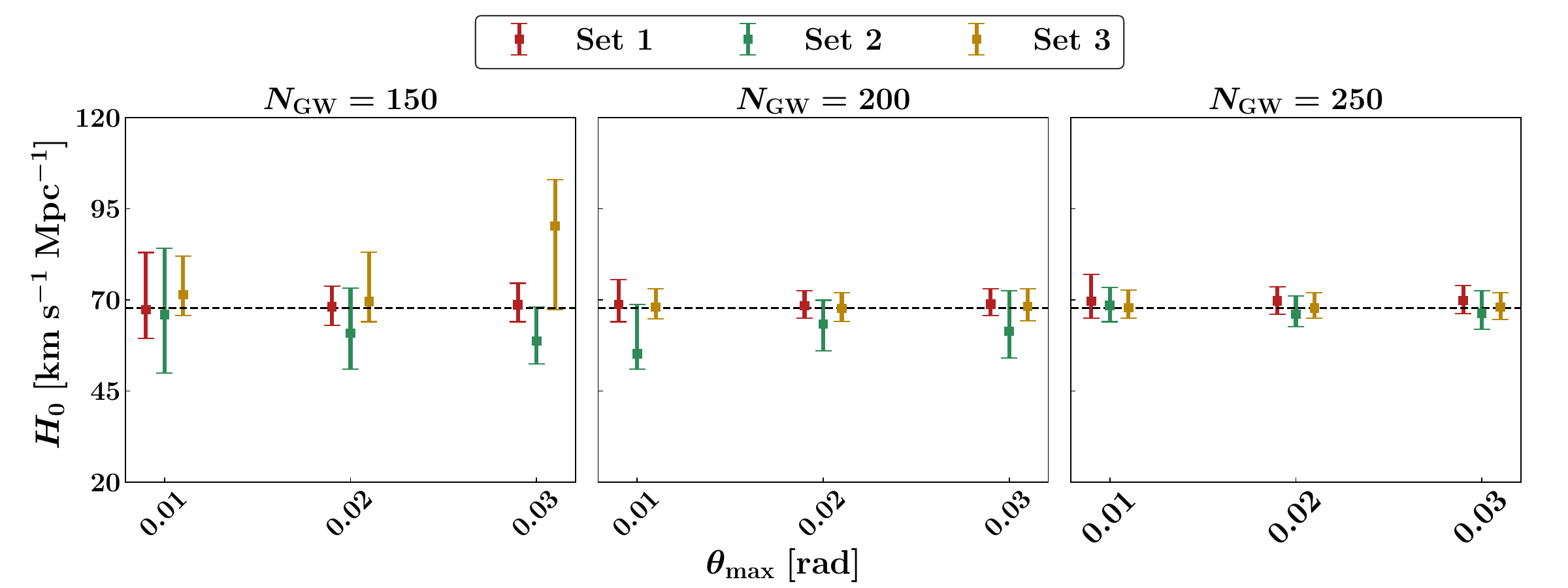}
    \caption{Uncertainty corresponding to $90\%$ highest density interval in estimating $H_{0}$ from different subsets of GW events for various values of $\theta_{\rm max}$ . The dashed horizontal line in each of the plots corresponds to the true value of $H_{0}=67.77~\kmsMpc$. Each plot corresponds to three different sets of a fixed number of GW events, mentioned at the top of the corresponding plot. $H_{0}$ error bar plot of set 2 GW events corresponds to the correct $\theta_{\rm max}$. However, $H_{0}$ error bar plot of set 1 and set 3 GW events are plotted with the offsets of $-0.001$ and $0.001$ radian in $\theta_{\rm max}$. These offsets are used only for illustration purposes.}
    \label{fig:H0error_compare}
\end{figure*}

We apply our method as mentioned in Sec.~\ref{sec:method} to three different mock GW catalogs for three different choices of $\theta_{\rm max} = \left\{0.01, 0.02, 0.03\right\}$ radian to study the statistical robustness. The GW catalogs are random realizations of the same population, as discussed in Sec.~\ref{sec:simulation}. Each GW catalog consists of $250$ simulated GW events that are detected with a network signal-to-noise ratio (SNR) $\rho_{\rm net} \geq 12$ in the three LIGO-Virgo detectors, with O4 design sensitivity~\cite{KAGRA:2013rdx}. We also study $H_{0}$ posteriors for two subsets of $150$ and $200$ GW events from those three sets of $250$ detected GW events. The threshold on network SNR in this work is chosen based on the earlier LVK cosmology papers~\cite{LIGOScientific:2019zcs, LIGOScientific:2021aug}. Sources with low network SNR would not contribute significantly to the improvement in the inference of the Hubble constant due to the poor sky localization and larger uncertainty in distance measurement. Figure~\ref{fig:H0_snr12} shows the $H_{0}$ posteriors for three different sets of GW events with varying numbers of detected events. Each row of Fig.~\ref{fig:H0_snr12} corresponds to a particular value of $\theta_{\rm max}$, mentioned in the right of the figure for different numbers of GW events. The columns of Fig.~\ref{fig:H0_snr12} correspond to $H_{0}$ posteriors for a fixed number of GW events, mentioned at the top of that panel, with varying $\theta_{\rm max}$. The black vertical dashed lines correspond to the injected value of $H_{0}=67.77~\kmsMpc$ in Fig.~\ref{fig:H0_snr12}.

We also summarize the uncertainty corresponding to $90\%$ highest density interval in constraining $H_{0}$ in Fig.~\ref{fig:H0error_compare} for a different number of detected GW events. The different panels correspond to varying numbers of detected events $N_{\rm GW}$.
In each panel, the error bar of $H_{0}$ uncertainty for three different sets is depicted with three different colors. Set $1$ and $3$ have offsets of $-0.001$ and $0.001$ radian, respectively, corresponding to the true $\theta_{\rm max}$ for clarity in the representation.
In Fig.~\ref{fig:H0error_compare}, the black horizontal dashed lines represent the injected value of $H_{0}=67.77~\kmsMpc$.

As expected, the posterior of $H_{0}$ starts to converge to the true value of $H_{0}=67.77~\kmsMpc$ with an increasing number of GW events. 
However, for the smaller number of GW events $\sim 150$, the $H_{0}$ posteriors can show a wide variation. We speculate that the non-Gaussian nature of galaxy overdensity fluctuation $\sigma_{\delta}$ (see Fig~\ref{fig:delobs_hist}) and not accounting for the covariance between different redshift bins is likely the reason for this wide variation. Though this is not an issue for unbiased estimation of $H_{0}$ with larger GW events, it is an aspect that needs to be carefully explored in the future. We have studied $H_{0}$ posteriors for larger $\theta_{\rm max} > 0.03$ radian and found the posteriors degrade even when the number of GW events is significant, i.e., $\gtrsim 150$. This is likely related to the fact that large values of $\theta_{\rm max}$ would decrease the amplitude of the correlation signal, reducing the total signal-to-noise ratio.

On the other hand, we have observed it becomes difficult to capture the correlation signal for smaller values of $\theta_{\rm max} \lesssim 0.005$ radian due to the significant fluctuation in measuring the clustering for a smaller number of galaxies, especially at low redshift. So, it is crucial to explore the optimal value of $\theta_{\rm max}$ to perform this method for unbiased estimation of the Hubble constant. From Fig.~\ref{fig:H0error_compare}, we observe that $\theta_{\rm max}=0.02$ radian may be the optimal choice in our study. However, the optimal value of $\theta_{\rm max}$ may vary for different GW events. In this work, we keep the values of $\theta_{\rm max}$ fixed for all GW events for simplicity. We leave the exercise of finding the optimal $\theta_{\rm max}$ given the localization area on the sky to investigations in the future.

\input{H0_measurement_table}

We report the $90\%$ highest density interval of the inferred values of $H_{0}$ corresponding to the three different sets in Table~\ref{tab:H0_measurement} for three different choices of $\theta_{\rm max}=\left\{0.01, 0.02, 0.03 \right\}$ radian. 
The corresponding $H_{0}$ posteriors are also shown in Fig.~\ref{fig:H0error_compare}.
We observe that we can constrain $H_{0}$ with a precision of $\lesssim 8\%$ ($90\%$ highest density interval) by combining the individual $H_{0}$ posteriors from $250$ GW events. In the best-case scenario, $H_{0}$ can be measured as precisely as $\sim 5\%$. 
For a smaller number of $150$ GW events, the uncertainty in measuring $H_{0}$ varies between $8\%-14\%$ across different sets of GW events when considering the optimal $\theta_{\rm max}=0.02$ radian. However, the uncertainty in $H_{0}$ increases to $\sim 25\%$ with other choices of $\theta_{\rm max}$. 
Following Ref.~\cite{Bera:2020jhx}, we also investigate the impact of our method on a smaller sample of nearby events, specifically $50$ events within $700$~Mpc, and present a representative example of the $H_{0}$ posterior for one of the realizations with a median accuracy of about $11\%$, in Fig.~\ref{fig:H0_rand50}. This precision is similar to that obtained in Ref.~\cite{Bera:2020jhx}, even after our careful accounting for the localization uncertainty. We speculate that the uncertainties degrade only slightly due to our improved treatment of the cross-correlation of galaxies with individual GW events, which compensates for the possible degradation due to localization uncertainty. However, we note that, with $50$ events, a minority of posteriors exhibit multimodal nature with a significantly increased uncertainty. We have cross-checked that by adding more events, the posterior converges to the correct mode.     
This suggests that measuring the Hubble constant from fewer events could lack statistical reliability, which could be remedied by increasing the sample of GW events to be as large as $250$.
It is important to note that we consider GW events within $1$ Gpc as the mock galaxy catalog used in this work is volume limited, extending up to redshift $z \sim 0.46$, already detailed in Sec.~\ref{sec:simulation}. 
The sky localizations of these simulated GW events are better than those from larger distances, which are observed by the current generation detectors. So, the broad sky localizations of GW events from significantly large distances may affect $H_{0}$ estimation and result in broadening the $H_{0}$ posterior.

\section{Discussion} \label{sec:summary}

Observations of GWs from coalescing binaries provide the direct measurement of the luminosity distance unaffected by systematics in various external astrophysical calibrations that rely on the cosmic distance ladder. However, GW estimates can be biased due to systematics inherent to its measurements, such as GW detector calibration~\cite{Sun:2021qcg, Bhattacharjee:2020yxe}. Assuming that the instrument can be calibrated sufficiently accurately~\cite{Sun:2021qcg, Bhattacharjee:2020yxe}, GW sources can be used to estimate the Hubble constant and other cosmological parameters. However, redshift information is not available from GW observations. Additionally, most of the detected GW events are BBH mergers and are not accompanied by EM counterparts. 
In this paper, we successfully demonstrated the inference of the Hubble constant from individual GW events by cross-correlating with the galaxy catalog. This method is useful even if the mergers do not have any EM observations and can supplement the inference of the Hubble constant from BNS mergers with detected EM counterparts.

In this work, we have considered varying numbers of GW events in the range of $[150, 250]$ to assess the performance of our method. The total number of GW events that we expect when we consider events from GWTC-3 as well as the ongoing fourth observation (O4) run of the LVK detectors\cite{KAGRA:2021vkt, KAGRA:2021duu} are similar. However, they will be distributed in a much larger volume than the $1$~Gpc in which we populate our simulated sources. We have restricted ourselves to the $1$~Gpc due to the limitations of the simulation volume available to us. The method that we show in our paper is, in principle, expected to be applicable to all GW events. The precision of the $H_{0}$ estimation depends on the population of GW events as the strength of the cross-correlation signal between GW events and galaxies is expected to diminish for GW events at larger distances due to the larger volume uncertainty. Given that the estimation of the Hubble constant (in particular, the credible intervals) is not robust (see Sec.~\ref{sec:results}), we refrain from analyzing the current GW events from GWTC-3 due to the small sample size. Once the new catalog of GW events, including those from O4, is available, we will explore the use of this method for the inference of the Hubble constant.

Our method consists of measuring the galaxy overdensity as a function of redshift at the maximum \textit{a posteriori} probability in the sky localization of individual gravitational wave events. We compare the data vector of observed galaxy overdensity $\bm{d}_{g}^{\rm obs}$ with that of the modeled galaxy overdensity $\bm{d}_{g}^{\rm mod}$ which depends on the value of $H_{0}$, to define a likelihood and infer the posterior distribution of $H_{0}$ in a Bayesian analysis. As examples, we considered three different mock GW catalogs to infer $H_{0}$ using the method described in Sec.~\ref{sec:method} and were able to constrain the Hubble constant with an accuracy of $\lesssim 8\%$ ($90\%$ highest density interval). The efficacy of this method is tested for the second-generation detector network comprised of the Advanced LIGO~\cite{LIGOScientific:2014pky} and Advanced Virgo~\cite{VIRGO:2014yos} detectors. With the improved sensitivity in the third-generation detectors like Cosmic Explorer~\cite{Evans:2021gyd} and Einstein Telescope~\cite{Maggiore:2019uih}, GW sources will be detected with much higher detection rate, and precise estimation of luminosity distance and sky localization~\cite{Borhanian:2022czq, Pieroni:2022bbh, Iacovelli:2022bbs, Branchesi:2023mws, Gupta:2023lga}. We expect this method to constrain $H_{0}$ with higher precision with such detectors.

In our current analysis, we have only considered $H_{0}$ as a free parameter to be constrained from the GW and galaxy overdensity observations. However, it is straightforward to extend our analysis to constrain other cosmological parameters, such as the matter density at the current epoch and dark energy equation of state. We do not consider them here to avoid the extra computational costs involved in estimating cosmological parameter(s). With the second-generation GW detectors, cosmological parameters except for $H_{0}$ are expected to be weakly constrained~\cite{LIGOScientific:2021aug}.

Our results show that the cross-correlation method can be utilized to infer $H_{0}$ from individuals from GW events as a proof of concept. We have made several simplifying assumptions in this work. The mock galaxy catalog used here is assumed to be volume limited with uniform depth over the entire sky and no limit on the galaxy fluxes. In contrast, real galaxy catalogs are flux-limited and also have varying depths in different directions. We defer the application of our method to galaxy catalogs, which include these imperfections in future work.
We have also considered the idealistic scenario where the GW events are the unbiased tracers of underlying matter distribution. In reality, the GW events and the galaxies trace the large-scale structure with a redshift-dependent bias~\cite{Vijaykumar:2020pzn, Libanore:2020fim}. This redshift-dependent bias could arise from the evolving way in which the gravitational wave events populate dark matter halos or due to selection effects. These effects need to be parametrized and marginalized to constrain the cosmological parameters from real data. We also have not accounted for any impact that weak gravitational lensing will have on the luminosity distance estimates due to the intervening large-scale structure.
Weak lensing induces nontrivial spatial correlations on the sky with galaxies at redshifts lower than that of the GW events, albeit with a smaller amplitude than the physical cross-correlation with galaxies at the same redshift as the GW events. This effect could be additionally incorporated into our methodology while working with real data, especially for high redshift events, by adding in a cross-correlation term related to weak lensing~\cite{Oguri:2016dgk}. Furthermore, weak lensing can also result in an increased covariance for the measurement of the overdensity. Our method of utilizing random lines of sight would naturally account for this increased covariance. The exact quantitative impact of such effects, however, will require full sky galaxy catalogs with weak lensing shears computed at multiple different lens planes and we defer such a study to future work.

\section*{Acknowledgments}

The authors acknowledge Divya Rana for helpful discussions throughout this work. The authors would like to thank Simone Mastrogiovanni, Aditya Vijaykumar, and Anuj Mishra for carefully reading the manuscript and useful suggestions. 
We also thank the anonymous referees for the careful review of the manuscript and the valuable comments.
T.G. acknowledges using the IUCAA LDG cluster Sarathi, accessed through the LVK collaboration, for the computation involved in this work. The work of S. Bera was supported by the Universitat de les Illes Balears (UIB); the Spanish Agencia Estatal de Investigación Grants No. PID2022-138626NB-I00, PID2019-106416GB-I00, RED2022-134204-E, RED2022-134411-T, funded by MCIN/AEI/10.13039/501100011033; the MCIN with funding from the European Union NextGenerationEU/PRTR (PRTR-C17.I1); Comunitat Autonòma de les Illes Balears through the Direcció General de Recerca, Innovació I Transformació Digital with funds from the Tourist Stay Tax Law (PDR2020/11 - ITS2017-006), the Conselleria d’Economia, Hisenda i Innovació grant numbers SINCO2022/18146 and SINCO2022/6719, cofinanced by the European Union and FEDER Operational Program 2021-2027 of the Balearic Islands. S. Bera received partial support from the French government under the France 2030 investment plan, as part of the Initiative d'Excellence d'Aix Marseille Université - A*MIDEX AMX-22-CEI-02. S. Bose acknowledges support from the NSF under Grant No. PHY-2309352.

The MultiDark Database used in this paper and the web application providing online access to it were constructed as part of the activities of the German Astrophysical Virtual Observatory as a result of a collaboration between the Leibniz-Institute for Astrophysics Potsdam (AIP) and the Spanish MultiDark Consolider Project No. CSD2009-00064. The  Big MultiDark Planck simulation has been performed in the Supermuc supercomputer at LRZ using time granted by PRACE.

\appendix

\section{DIFFERENCES AND IMPROVEMENTS OVER THE PREVIOUS WORK}\label{sec:difference}

A general overview of the differences between this study and Ref.\cite{Bera:2020jhx}, both of which utilize the clustering of GW events with galaxies, pertaining to their methodology has been highlighted at the end of Sec.\ref{sec:method}. Below, we provide more details comparing the two studies with regard to the simulated GW source population as well as the analysis technique.

To test the \textit{clustering} method, in Ref.~\cite{Bera:2020jhx}, the authors used simulated GW events up to a distance $ \sim 1300$~Mpc (or, $900h^{-1}$~Mpc). They divided the GW events into bins of width $200$ Mpc, based on the median luminosity distance of the posterior distribution of the luminosity distance. For a sample of GW events in a given luminosity distance bin, they then proceeded to compute the angular cross-correlation with galaxies [see Eq.~(13) in Ref.~\cite{Bera:2020jhx} for details on the estimator of angular cross-correlation] using the maximum probable posterior sky position of each GW event. The uncertainties in the angular position of the GW event are expected to reduce the amplitude of the cross-correlation signal, but this was never modeled in Ref.~\cite{Bera:2020jhx}. Instead, they utilized the fact that the angular cross-correlation for each distance bin would still peak at the correct redshift corresponding to the GW events in the luminosity distance bin and modeled it as a Gaussian distribution [Eq. (14) in Ref.~\cite{Bera:2020jhx}], to estimate the mean redshift and its error. For each bin, they assigned an error-weighted average luminosity distance based on the errors in the luminosity distance from the posterior distribution of parameters given the GW data. 
The resultant Hubble diagram (luminosity distance versus redshift) was then used to estimate $H_{0}$. Their study thus used a variety of simplifications to demonstrate a proof of concept of how the clustering of gravitational waves with galaxies can be used to infer the Hubble constant.

In this work, we incorporate several further improvements. In contrast with the earlier work, we adopt a single-step Bayesian inference method, eliminating the need to bin the GW events into luminosity distance bins. Instead of measuring the angular cross-correlation of a sample of GW events with galaxies, we measure the overdensity of galaxies around each GW event separately. The biggest difference is in our modeling scheme. We model the $3$D cross-correlation of GW events with galaxies, thus predicting both the amplitude and shape of how the overdensities around GW events depend upon the cosmological parameters. In our work, we also account for the posterior distribution of the sky localization as well as the luminosity distance while modeling the observables, thus naturally accounting for the smearing of the angular cross-correlation function. In this manner, any non-Gaussianities in the posterior distributions, as well as the correlated errors in the localization volume, are automatically incorporated in our study.

Finally, we also comment on the precision of the inference of the Hubble constant obtained in the two studies. In the earlier work, a large number of GW events were distributed between $200$ and $1300$~Mpc, out of which $50$ and $500$ (distributed up to $\sim 700$~Mpc) sources were selected to infer the Hubble constant with an accuracy of $\sim 10\%$ and $2\%$, respectively. In the present work, in addition to analyzing 250 events out to $1$ Gpc, we have also repeated the exercise for $50$ events and shown in the main text (see Fig.~\ref{fig:H0_rand50} ) that we get similar accuracy for the inference on $H_0$ as in the earlier study even after marginalizing over the localization uncertainty.

\bibliography{reference}

\end{document}

%% file: H0_measurement_table.tex
\begingroup
\begin{table}
\begin{center}
\begin{tabular}{|c | c | c@{\hskip 1.1em} c@{\hskip 1.1em} c |}
\hline
\multirow{2}{*}{\centering $N_{\rm GW}$} & \multirow{2}{*}{\centering $\theta_{\rm max}$ (radian)} & \multicolumn{3}{c|}{\centering $H_{0} ($\kmsMpc$)$}\\
\cline{3-5}
& & Set $1$ & Set $2$ & Set $3$ \\
\hline
\multirow{3}{*}{$150$} & $0.01$ & $67.30^{+15.69}_{-7.84}$ & $65.97^{+18.19}_{-15.97}$ & $71.41^{+10.59}_{-5.77}$\\
                       & $0.02$ & $68.10^{+5.74}_{-5.10}$ & $60.90^{+12.26}_{-9.90}$ & $69.57^{+13.50}_{-5.57}$\\
                       & $0.03$ & $68.65^{+5.91}_{-4.64}$ & $58.73^{+9.27}_{-6.29}$ & $90.28^{+12.72}_{-23.01}$\\ [0.5ex]
\hline
\multirow{3}{*}{$200$} & $0.01$ & $68.76^{+6.71}_{-4.76}$ & $55.24^{+13.59}_{-4.24}$ & $68.02^{+4.97}_{-3.19}$\\
                       & $0.02$ & $68.37^{+4.09}_{-3.37}$ & $63.34^{+6.58}_{-7.34}$ & $67.64^{+4.35}_{-3.51}$ \\
                       & $0.03$ & $68.85^{+4.14}_{-3.23}$ & $61.42^{+11.07}_{-7.42}$ & $68.22^{+4.78}_{-3.90}$\\ [0.5ex] 
\hline 
\multirow{3}{*}{$250$} & $0.01$ & $69.66^{+7.41}_{-4.67}$ & $68.53^{+4.92}_{-4.53}$ & $67.82^{+4.84}_{-2.82}$\\
                       & $0.02$ & $69.68^{+3.83}_{-3.68}$ & $66.07^{+4.93}_{-3.48}$ & $66.07^{+4.92}_{-3.48}$\\
                       & $0.03$ & $69.84^{+4.15}_{-3.54}$ &  $66.30^{+6.18}_{-4.30}$ & $68.00^{+4.00}_{-3.34}$ \\ [0.5ex] 
\hline
\end{tabular}
\caption{\label{tab:H0_measurement}Summary of uncertainty (maximum \textit{a posteriori} probability with $90\%$ highest density interval) of inferred values of the Hubble constant from different sets of GW events with varying $\theta_{\rm max}$.}
\end{center}
\end{table}
\endgroup